\documentclass{article}

\usepackage{amsmath, amssymb, latexsym}

\newtheorem{thm}{\bf Theorem}[section]
\newtheorem{theorem}{\bf Theorem}[section]
\newtheorem{lemma}[thm]{\bf Lemma}        
\newtheorem{proposition}[thm]{\bf Proposition}  

\newtheorem{corollary}[thm]{\bf Corollary}

\newtheorem{definition}[thm]{\bf Definition}

\newtheorem{example}[thm]{\bf Example}

\newtheorem{remark}[thm]{\bf Remark}

\newcommand{\Bbox}{
{\unskip\nobreak\hfil\penalty50
\hskip1em\hbox{}\nobreak\hfil{\lower .5pt \hbox{$\Box$}}
\parfillskip=0pt \finalhyphendemerits=0 \par}
}

\newcommand{\eop}{
\ifmmode {\hbox{\Bbox}} \else \Bbox \fi
}

\newcommand{\x}{\times}

\begin{document}
\newcommand{\LO}{\textbf{LO}}
\newcommand{\Fr}{\textsf{Fr}} 
\newcommand{\Lfr}{\textsf{Lfr}} 
\newcommand{\dom}{\textsf{dom}}
\newcommand{\F}{\mathcal{F}}
\newcommand{\C}{\mathcal{C}}
\newcommand{\tos}{T^{\omega}_\Sigma}
\newcommand{\ooo}{\omega ^{\omega^\omega }}
\newcommand{\bin}{\mathbf{bin}}
\newcommand{\hgt}{\mathbf{ht}}
\renewcommand{\o}{\mathbf{o}}
\newcommand{\CFL}{\textsl{CFO}} 
\newcommand{\reg}{\textbf{reg}}
\newcommand{\cf}{\textbf{cf}} 
\newcommand{\tot}{\textbf{tot}}
\newcommand{\wo}{\textbf{wo}} 
\newcommand{\sca}{\textbf{sca}}
\newcommand{\den}{\textbf{den}} 
\newcommand{\RL}{\mathbf{RL}} 
\newcommand{\ol}{\overline}
\newcommand{\cA}[1]{\hbox{${\cal A}_{#1}$}} 
\newcommand{\ax}{\hbox{\sl Ax}}
\newcommand{\1}{\mathbf{1}}
\newcommand{\zos}{\{0,1\}^*}
\newcommand{\zop}{\{0,1\}^+}
\newcommand{\lex}{<_\ell}
\newcommand{\slex}{<_\ell}
\newcommand{\Q}{\mathbb{Q}}
\newcommand{\op}{{\omega ^{op}}}
\renewcommand{\r}{ ^{\small op}}
\renewcommand{\S}{\hbox{${\cal S}$}}
\renewcommand{\L}{\hbox{${\cal L}$}}
\renewcommand{\P}{\hbox{${\cal P}$}}
\newcommand{\G}{\hbox{${\cal G}$}}
\newcommand{\N}{\mathbb{N}}
\newcommand{\Z}{\mathbb{Z}}
\newcommand{\sqleq}{\sqsubseteq}
\newcommand{\sqle}{\sqsubset}

\newcommand{\dv}{\stackrel{*}{\implies}} 
\newcommand{\odv}{\stackrel{\circledast}{\implies}}

\title{Algebraic Ordinals\footnote{This is a revised version of the paper submitted July 5, 2009}} 
\author{
S. L. Bloom\thanks{Supported in part by
the TAMOP-4.2.2/08/1/2008-0008 program of the Hungarian National Development Agency.} \\
Department of Computer Science\\
Stevens Institute of Technology\\
Hoboken, NJ, USA\\
\and
Z. \'Esik\thanks{Supported in part by grant no. K 75249 from the 
National Foundation of Hungary for Scientific Research and by
the TAMOP-4.2.2/08/1/2008-0008 program of the Hungarian National Development Agency.
} \\
Department of Informatics \\
University of Szeged\\
Szeged, Hungary}


\maketitle

\begin{abstract}
An \textbf{algebraic tree} 
 $T$ is one determined by a finite system of fixed point equations.
The frontier $\Fr(T)$ of an algebraic tree $T$ is linearly ordered by the lexicographic order
$\lex$. If {$(\Fr(T)\lex)$}  is well-ordered, its order type
is an \textbf{algebraic ordinal}.
We prove that the algebraic ordinals are exactly the ordinals less than $\ooo$.
\end{abstract}


\section{Introduction}\label{ }

Fixed points and finite systems of fixed point equations occur in just about 
all areas of computer science. Regular and context-free languages, 
rational and algebraic formal power series, finite state process 
behaviors can all be characterized as (components of) canonical solutions 
(e.g.,  unique, least or greatest, or initial solutions) 
of systems of fixed point equations.

Consider the fixed point equation
\begin{eqnarray*} 
X &=& 1 + X
\end{eqnarray*} 
over linear orders, where $+$ denotes the sum operation (functor)
on linear orders. As explained in \cite{BEMezei},
its canonical (initial) solution is the ordinal $\omega$, 
or any linear order isomorphic to 
the ordering of the natural numbers.
For another example, consider the system of fixed point equations 
\begin{eqnarray*}
X &=& Y + X\\
Y &=& 1 + Y
\end{eqnarray*}
The first component of its canonical solution is $\omega^2$,
and the second component is $\omega$. Of course, there exist
fixed point equations whose canonical solution is not well-ordered, 
for example, the canonical solution of 
\begin{eqnarray*}
X &=& X + 1
\end{eqnarray*} 
is $\omega^*$, the reverse of $\omega$, and the canonical solution of  
\begin{eqnarray*}
X &=& X + 1 + X 
\end{eqnarray*} 
is the ordered set of the rationals.

The above equations are quite simple since they involve no parameters. 
The unknowns $X,Y$ range over linear orders, or equivalently, 
constant functions (or rather, functors) defined on linear orders.  By allowing unknowns 
ranging over functions (or functors) in several variables, we obtain 
the ordinal $\omega^\omega$ as the first component of the canonical solution of 
\begin{eqnarray*}
F_0 &=& G(1)\\
G(x) &=& x + G(F(x))\\
F(x) &=& x + F(x)
\end{eqnarray*} 
The second and third components of the canonical solution are functors 
which map a linear order $x$ to $x \times \omega^2$ and $x \times \omega$, 
respectively.

We call a linear order algebraic if it is isomorphic to 
the first (or principal) 
component of the canonical solution of a system of fixed point 
equations of the sort 
\begin{eqnarray}
\label{eq-sys}
F_i(x_0,\ldots,x_{n_i-1}) &=& t_i,\quad i = 1,\ldots, n
\end{eqnarray}
where $n_1 = 0$ and each $t_i$ is an expression composed of the function variables 
$F_j$, $j = 1,\ldots, n$, the individual variables $x_0,\ldots,x_{n_i-1}$, 
the constant $1$ and the sum operation $+$. Moreover, we call a linear 
order regular if it is isomorphic to the first component of the canonical solution 
of a system (\ref{eq-sys}) with $n_i = 0$ for all $i$. 
Further, we call an ordinal algebraic or regular if it is an algebraic 
or regular linear order. 

From the results in \cite{Heilbrunner}, it follows easily that
an ordinal is regular if and only if it is less than $\omega^\omega$ 
(See also \cite{BlCho01}.)
It was conjectured in \cite{BEbergen} that an ordinal is algebraic if and only if 
it is less than $ \omega^{\omega^\omega}$. Our aim in this paper is to 
confirm this conjecture. 

Infinite structures may 
be described by finite presentations in several 
different ways. One method, represented by automatic structures, consists in 
describing a structure up to isomorphism by representing its 
elements as words or trees or some other combinatorial objects, 
and its relations by rewriting rules or automata. Another, algebraic approach 
is describing an infinite structure as the canonical solution of a 
system of fixed point equations over a suitably defined algebra of structures. 
Our results show that these methods are equivalent, at least for 
small ordinals.\footnote{For another result in this direction, see 
\cite{Colcombet}.} 

An automatic relational structure \cite{Hodsgon,KhoussainovNerode} is a countable 
structure whose carrier is given by a regular language and whose relations 
can be computed by synchronous multi-tape automata. A tree automatic 
structure \cite{DauchetTison,Delhomme,Colcombet} is defined using tree automata,
cf. \cite{GS}. It was proved in \cite{Delhomme} that the automatic ordinals are 
exactly the ordinals less than $\omega^\omega$. See also \cite{Khoussainovetal}. 
In \cite{Delhomme}, it is also shown that an ordinal is tree automatic
if and only if it is less than $ \omega^{\omega^\omega}$.  
Thus, an ordinal is automatic if and only if it is regular\footnote{Compare this 
fact with Theorem 2 in \cite{Broughetal}.}, and is tree 
automatic if and only if it is algebraic. Actually the claim that every 
regular ordinal is automatic is immediate. It would be interesting to 
derive a direct proof of the fact that every algebraic ordinal is 
tree automatic, or the other way around.

In a traditional setting, one solves a system of fixed point equations
in an algebra equipped with a suitable partial order; 
there is a least element, suprema of $\omega$-chains
exist, the operations preserve the ordering 
and least upper bounds of $\omega$-chains.  Such algebras
are commonly called continuous algebras (or $\omega$-continuous algebras), cf. \cite{ADJ,Guessarian}. 
In this setting, one solution of this kind of system is provided
by least fixed points.  The classical Mezei-Wright theorem \cite{Mezei-Wright} asserts that such
a solution is preserved by a continuous, order preserving algebra 
homomorphism.

However, in several settings such as (countable) linear orders, 
there is no well-defined partial order but one can naturally 
introduce a category by considering morphisms between linear orders.
A generalization of the classical Mezei-Wright theorem  
to the setting of  ``continuous categorical algebras'' 
has been given in \cite{BEMezei},
where least 
elements are replaced by initial elements,
and suprema of $\omega$-chains are replaced by colimits of $\omega $-diagrams.
Since trees, equipped with the usual partial order 
\cite{ADJ,Guessarian,BEMezei} form an initial continuous categorical algebra, 
it follows that instead of solving a system of fixed point equations 
directly over linear orders, we may first find its least solution over trees, 
to obtain an algebraic or regular tree, and then take the image 
of this tree with respect to the unique morphism from trees to linear 
orders which assigns 
to a tree the linear order determined by the frontier 
of the tree.  Thus, up to isomorphism, an algebraic (or regular) linear order 
is the frontier of an algebraic (or regular) tree. In this way, we may represent 
algebraic and regular linear orders and ordinals as frontiers of algebraic
or regular trees, and this is the approach we take here. 
This approach is not new. Courcelle was the first to use 
frontiers of trees to represent linear orders and words, i.e., labeled 
linear orders. The origins of the notions of regular linear order and regular word
go back to \cite{Courcelle78}.  

In our argument showing that every algebraic ordinal is less than
$\omega^{\omega^\omega}$, we will make use of certain context-free
grammars generating prefix languages, called ``ordinal grammars''
which seem to have independent interest. 
Equipped with the lexicographic order (see below), the language 
generated by an ordinal grammar is well-ordered. We show that 
an ordinal is the order type of a language generated by an 
ordinal grammar if and only if it is less than $\omega^{\omega^\omega}$.
We then show how to translate an algebraic tree given by a 
system of equations to an ordinal grammar generating the frontier of the tree. 

The paper is organized as follows. In Section~\ref{sec-basics},
we define regular and algebraic trees and linear orders, and 
regular and algebraic ordinals. Then, in Section~\ref{sec-closure},
we establish some closure properties of algebraic linear orders, 
including closure under sum, multiplication, and $\omega$-power,
and use these closure properties to establish that any ordinal 
less than $ \omega^{\omega^\omega}$ is algebraic. Section~\ref{sec-grammars}
is devoted to ordinal grammars and the proof of the result that 
an ordinal is the 
order type 
of the context-free language generated by an ordinal 
grammar, equipped with the lexicographic order, 
if and only if it is less than $ \omega^{\omega^\omega}$. Then, in Section~\ref{sec-translation}
we show how to construct for an algebraic tree $T$ (given by a system of 
fixed point equations) an ordinal grammar $G$ such that the order type of the 
language generated by $G$ equals the order type of the frontier of $T$. 
(The proof of the correctness of the translations is moved to an appendix.)
The paper ends with  Section~\ref{sec-conclude} containing some concluding remarks.

\section{Linear orders, words, prefix languages, and trees} 
\label{sec-basics}

\subsection{Linear orders}  

A linearly ordered set $(A,<_A)$ consists
of a set $A$ equipped with a strict linear order, i.e., an irreflexive and 
transitive binary relation $<_A$ 
that satisfies exactly one of the conditions
\begin{eqnarray*}
  x = y\ \text{ or }\ x<_A y \text{ or }\ y<_A x,
\end{eqnarray*}
for all $x,y \in A$.  If $(A,<_A)$ and $(B,<_B)$ are linearly ordered sets,
a \textbf{morphism} $h:(A,<_A) \to (B,<_B)$ is a function $h:A \to B$
such that for all $x,y \in A$, if $x<_A y $ then $h(x)<_B h(y)$.  The collection of all linearly
ordered sets and morphisms forms a category \LO.  We say two linearly ordered
sets are isomorphic if they are isomorphic in \LO.  
The \textbf{order type} of a linear order is the isomorphism class 
of the linear order in \LO.  We write $\o(A,<_A)$ for the order type of $(A,<_A)$.

A linearly ordered set $(A,<_A)$ is \textbf{well-ordered} if 
every nonempty subset of $A$ has a least element.
If $\alpha $ is an ordinal, then $\alpha $ is well-ordered
by the membership relation $\in $. 
When $(A,<_A)$ is well-ordered, there is a unique ordinal 
$\alpha $ such that $(\alpha, \in)$ is isomorphic to $(A,<_A)$, 
and the order type of $(A,<_A)$ is conveniently identified with this ordinal.

\textbf{Remark}.
We remind the reader that if $\alpha =\o(A,<_A)$ and $\beta =\o(B,<_B)$ are
ordinals, where $A,B$ are disjoint sets, then $\alpha + \beta $ is the
order type 
of $A \cup B$, ordered by putting every element of $A$ before
every element of $B$; otherwise, imposing the given ordering on elements inside
$A$ or $B$. Similarly, if $\alpha_n=\o(A_n,<_n)$ is an ordinal for each $n \geq 0$, 
where the sets $A_n$ are pairwise disjoint, 
then $\sum_n \alpha_n$ is the order type of the union $\bigcup_n A_n$ ordered by
making every element of $A_n$ less than every element in $A_m$, for $n<m$, and imposing
the given order on $A_n$.

The ordinal $\alpha \x \beta $ is the order type of $A \x B$, ordered by ``last differences'',
i.e.,  where 
\begin{eqnarray*}
  (a,b) < (a',b') &\iff & b <_B b' \text{ or } (b=b' \text{ and  } a <_A a').
\end{eqnarray*}

Last, $\alpha ^ \omega $ is the least upper bound of all ordinals $\alpha ^n = 
\overbrace{\alpha \x \ldots \x \alpha }^n $, for
$n < \omega $.

For more definitions and facts on linear orders and ordinals 
we refer to \cite{Roitman,Rosenstein,Sierpinski}. 
For later use, we prove:

\begin{lemma}
\label{lem-ordinals}
Suppose that $\alpha$ is an ordinal and $\beta$ is an infinite ordinal,
so that $1 + \beta = \beta$. If $\alpha$ is a successor ordinal then 
\begin{eqnarray*}
(\beta + 1)\times \alpha  &\leq & \beta \times \alpha + 1.
\end{eqnarray*} 
If $\alpha$ is $0$ or a limit ordinal, then 
\begin{eqnarray*}
(\beta + 1) \times \alpha &\leq & \beta \times \alpha.
\end{eqnarray*}
\end{lemma} 

{\sl Proof.\/} We prove both claims by transfinite induction on $\alpha$. 
The case when $\alpha = 0$ is clear. Assume that $\alpha$ is 
a successor ordinal. Then $\alpha = \gamma + n$ for some 
positive integer $n$ where $\gamma$ is $0$ or a limit ordinal. 
Then, using the induction hypothesis and the assumption $1 + \beta = \beta$,  
\begin{eqnarray*}
(\beta + 1) \times \alpha &=& (\beta + 1)\times \gamma + (\beta + 1)\times n\\
&\leq & \beta \times \gamma + \beta \times n + 1\\
&=& \beta \times (\gamma + n) + 1 \\
&=& \beta \times \alpha + 1. 
\end{eqnarray*} 
Suppose now that $\alpha$ is a limit ordinal. Then,
\begin{eqnarray*}
(\beta + 1)\times \alpha &=&
\sup_{\gamma < \alpha}((\beta + 1)\times \gamma)\\
&\leq & 
\sup_{\gamma < \alpha} (\beta \times \gamma + 1)\\
&\leq& \beta \times \alpha,
\end{eqnarray*}
again by the induction hypothesis. \eop

\subsection{Words}  
We use $\N$ for the set of nonnegative integers.  If $n \in \N$, then 
$$[n]=\{0, \ldots,n-1\},$$
so that $[0] = \emptyset$.
The collection of all finite words on a set $A$ is denoted $A^*$.
The empty word is denoted $\epsilon$.
The set of nonempty words on $A$ is $A^+ = A^*\setminus\{\epsilon\}$. 
 We denote the product (concatenation)
of two words $u,v \in A^* $ by $uv$. We identify an element $a \in A$ with
the corresponding word of length one.

When $A$ is linearly ordered by $<_A$, then
$A^*$ is equipped with two partial orders.  The \textbf{prefix order}, written $u <_p v$,
is defined by:
\begin{eqnarray*}
  u <_p v &\iff & v = uw,\quad w \neq \epsilon ,
\end{eqnarray*}
for some word $w$.  The \textbf{strict} or \textbf{branching order},  
written $u<_s v$ is defined by:
$  u <_s v $ if and only if
 for some words $u_1, u_2, v_2$, and $a,b \in A$,
\begin{eqnarray*}
 u &=& u_1 a u_2\\
  v &=& u_1 b v_2, \quad  \text{ and  } \\ 
  a &<_A & b.
\end{eqnarray*}
The \textbf{lexicographic order} on $A^*$ is defined by:
\begin{eqnarray*}
  u \lex v &\iff & u<_p v \text{ or } u <_s v.
\end{eqnarray*}
It is clear that for any $u,v \in A^*$, exactly one of the following possibilities holds:
\begin{eqnarray*}
  u=v,\ u <_p v,\ v <_p u, \ u<_s v,\ v<_s u.
\end{eqnarray*}
Thus, the lexicographic order is a linear order on $A^*$.

\subsection{Prefix languages}

A \textbf{prefix language} on a set $A$ 
is a subset $L$ of $ A^*$ such that
if $u \in L$ and $uv \in L$ then $v=\epsilon $.
For later use, we state the following facts about prefix languages.
\begin{lemma}
\label{pref lemma}
\begin{itemize}
\item If $(P,<_P)$ is a countable linearly ordered set, there is a prefix language
$L$ on $[2]$ such that $(P,<_P)$ and $(L,\lex)$ are isomorphic. 
\item If $L$ is a prefix language, then, for $u,v \in L$,  $u \lex v \iff u<_s v$.
\item 
If $K,L \subseteq A^*$ are prefix languages, where $A$ is linearly ordered, 
so is $KL=\{uv:u \in K, v \in L \} $ and, when $(K,\lex)$ and $(L,\lex)$ are well-ordered,
  \begin{eqnarray*}
    \o( KL, \lex) &=& \o(L,\lex) \x \o(K,\lex).
  \end{eqnarray*}
\item 
Suppose that $L \subseteq [2]^*$ is a prefix language such that $\alpha = \o(L,\lex)$ is an
ordinal.
Then
\begin{eqnarray*}
(\bigcup_{n \geq 0} 1^n 0 L^n,\lex)
\end{eqnarray*}
is well-ordered and, if $\alpha > 1$, 
\begin{eqnarray}
\label{omega pow}
\alpha^\omega &=& \o(\bigcup_{n \geq 0} 1^n 0 L^n,\lex).
\end{eqnarray}
\end{itemize}
\end{lemma}
{\sl Proof.\/ } 
For the first statement we refer to \cite{Courcelle78,BEs}. 
We prove only the statement (\ref{omega pow}).  
The set, say $L_\infty $, of words on the right side of
(\ref{omega pow}) is the set of all words of the form
$1^n 0 u$, for $u \in L^n$. For any words $u,v$, if $1^n 0 u <_p 1^m 0 v$,
then $n = m$ and $u <_p v$. Since $L$ is a prefix language, so is $L^n$ and, thus
so is $\bigcup_n 1^n 0 L^n$. Also, if $n<m$, $1^n 0 u <_s 1^m 0 v$, and
$1^n 0 u <_s 1^n 0 v$ if and only if $u<_s v$. Thus, by definition, the order type of the 
lexicographic order
of the prefix language $L_\infty $ is $\sum_n \alpha^n$. It is now easy to
see that this sum is $\alpha^\omega $. \eop 

\subsection{Trees}  
Suppose that $\Sigma $ is a (finite) 
ranked alphabet, i.e., a nonempty finite set partitioned
into subsets $\Sigma_k$ of ``$k$-ary operation symbols'', $k\geq 0$.  
Moreover, suppose that $V = \{x_0,x_1,\ldots\}$ is an ordered set of 
``individual variables''. 
A \textbf{$\Sigma $-tree} $T$ in the variables $V$, or a 
``tree over $\Sigma$ in the individual variables $V$'', is a partial 
function $[n]^* \to \Sigma \cup V$
satisfying the conditions listed below.  Here, $n$ is largest such that
$\Sigma_n \neq \emptyset  $. The conditions are:
\begin{itemize}
\item The domain of $T$, $\dom(T)$, 
 is prefix closed: if $T(uv)$ is defined, then
so is $T(u)$.
\item If $T(u) \in \Sigma_k$, $k>0$, and  $T(ui)$ is defined, then  $i \in  [k]$.
\item If $T(u) \in \Sigma_0 \cup V$, then $u$ is a leaf, and $T(ui)$ is undefined, for
all $i$.
\end{itemize}
A $\Sigma$-tree $T$ is \textbf{complete} if whenever $T(u)$ is defined 
in $\Sigma_n$, for some $u$, then $T(u0),\ldots,T(u(n-1))$ are all defined.  Moreover,
$T$ is \textbf{finite} if its domain is finite. Below we will
usually denote finite trees by lower case letters. 
The size of a finite tree $t$ is the size of the set $\dom(t)$.   
The set of all $\Sigma $-trees in the variables $V$ is denoted $\tos(V)$.
Moreover, for a subset $V_n = \{x_0,\ldots,x_{n-1}\}$ of $V$, we write $\tos(V_n)$ 
for the collection of all trees all whose leaves are labeled in $\Sigma_0 \cup V_n$. 
When $n = 0$, we write simply $\tos$.

Trees are equipped with the following partial order $T \sqle T'$: 
Given $T,T' \in \tos(V)$ such that $T \neq T'$, we define 
$T \sqle T'$ if and only if for all words $u$, if $T(u)$ is defined, then $T(u) = T'(u)$. 
It is well-known that the partially ordered set $(\tos(V), \sqle)$ is $\omega $-complete,
i.e., $\tos(V)$ has as least element $\bot$, the totally 
undefined tree, 
and least upper bounds of all $\omega$-chains. Indeed, if $T_0 \sqle T_1\sqle \ldots $
is an $\omega $-chain in $\tos(V)$, the supremum 
 is the tree $T$ whose domain is the union
of the domains of the trees $T_n$, and, if $u $ is a word in this union,
$T(u)=\sigma \in \Sigma$ if and only if $T_n(u)=\sigma $, for some $n$. 
Similarly, $T(u) = v_i$ for some $v_i \in V$ if and only if $T_n(u)=v_i$ for some $n$.
Each symbol $\sigma \in \Sigma_k$ 
induces a $k$-ary operation on $T_\Sigma^\omega(V)$ in the usual way. 
It is well-known that these operations are $\omega$-continuous in all arguments
and $T_\Sigma^\omega(V)$ is an $\omega$-continuous algebra.
In the same way, for every $n$, $\tos(V_n)$ is an $\omega$-continuous algebra
(in fact, the free
$\omega$-continuous $\Sigma$-algebra on $V_n$). 
See \cite{ADJ,Guessarian}.

\subsection{Tree substitution}

In this section we define a substitution operation on trees, sometimes 
called second-order substitution.

Suppose that $\Sigma$ and $\Delta$ are ranked alphabets 
and for each $\sigma \in \Sigma_n$ we are given 
a tree $R_\sigma \in T^\omega_\Delta(V_n)$. 
We define substitution in two steps, first for finite trees, 
and by continuity for infinite trees. 
For each finite tree $t \in T_\Sigma(V)$ we define the tree $S = t[\sigma \mapsto R_\sigma]_{\sigma \in \Sigma}$
in $T^\omega_\Delta(V)$, sometimes denoted just $t[\sigma \mapsto R_\sigma]$
by induction on the size of $t$. When $t$ is the empty tree $\bot$, 
so is $S$. When $t$ is $x$, for some $x \in V$, then $S = x$.
Otherwise $t$ is of the form $\sigma(t_0,\ldots,t_{n-1})$, 
where $\sigma \in \Sigma_n$, 
and we define 
$$S = R_\sigma(t_1',\ldots,t_n')$$
where $t_i' = t_i[\sigma \mapsto R_\sigma]$, for all $i$. 

Suppose now that $T$ is an infinite tree in $T_\Sigma(V)$. Then there is an 
ascending $\omega$-chain $(t_n)$ of finite trees such that 
$T = \sup_n t_n$. We define 
$$T[\sigma \mapsto R_\sigma] = \sup_n t_n[\sigma \mapsto R_\sigma].$$

It is known, see \cite{CourcelleFund}, that substitution is $\omega$-continuous. 

\begin{proposition}
\label{proposition-subst}
Substitution is a continuous function
$$T^\omega_\Sigma(V) \times \prod_n  T^\omega_\Delta(V_n)^{\Sigma_n} \to T^\omega_\Delta(V).$$
\end{proposition}

Below when $\Sigma$ and $\Delta$ are not disjoint and $R_\sigma = \sigma(x_1,\ldots,x_n)$ 
for some $\sigma \in \Sigma_n$, then we often omit $\sigma$ from the arguments of the substitution.

\subsection{Algebraic trees and ordinals}
  
Now consider a finite system $E$ of equations of the form
\begin{eqnarray}
\label{alg tree sys}
F_1(x_0,\ldots,x_{n_1-1}) &=& t_1(x_0,\ldots,x_{n_1-1}) \\
\nonumber
F_2(x_0,\ldots,x_{n_2-1}) &=& t_2(x_0,\ldots,x_{n_2-1}) \\
\nonumber
  & \vdots & \\
\nonumber
F_m(x_0,\ldots,x_{n_m-1}) &=& t_m(x_0,\ldots,x_{n_m-1}),
\end{eqnarray}
where, for $i =1,\ldots,m$, $t_i$ is a \textbf{term} 
over the ranked alphabet $\Sigma \cup \F$ in the variables $\{x_0,\ldots,x_{n_i-1}\}$ 
 (i.e., finite complete tree in $T_\Sigma^\omega(V_{n_i})$), 
where $\F=\{F_1,\ldots,F_m\}$ is the set of ``function variables''
and each $F_i$ has rank $n_i$.  
See the example in (\ref{eq:example3}) below.
 Each term $t_i$ 
 induces a function
$$t_i^E: T^\omega_\Sigma(V_{n_1})\times \ldots \times  T^\omega_\Sigma(V_{n_m})
\to T^\omega_\Sigma(V_{n_i})$$
by substitution: 
$$(R_1,\ldots,R_m) \mapsto t_i[F_j \mapsto R_j]_{1 \leq j \leq m}.\footnote{It is 
understood here that each $\sigma \in \Sigma_k$ remains unchanged, i.e., gets substituted by 
$\sigma(x_0,\ldots,x_{k-1})$.}$$
By Proposition~\ref{proposition-subst} this function is $\omega$-continuous. 
The target tupling 
$$\langle t_1^E,\ldots,t_m^E\rangle$$
mapping  
$T^\omega_\Sigma(V_{n_1})\times \ldots \times  T^\omega_\Sigma(V_{n_m})$ 
to itself is also $\omega$-continuous and has a least fixed point 
$(T_1,\ldots,T_m)$.  
Thus, there is a least 
solution $(T_1,\ldots,T_m)$ 
of any such system.  One function
variable $F_i$  of rank 0   
is selected as the \textbf{principal variable}, and
the corresponding tree $T_i$ 
is the \textbf{principal component} of the least solution.
(Typically, we choose the first function variable as the principal variable.)
If every integer $n_i$, $i = 1,\ldots,m$, is zero, the system is said to be
\textbf{regular}. 

\begin{definition}
A tree $T$ in $\tos(V_k)$ is \textbf{algebraic in $T^\omega_\Sigma(V_k)$},
(respectively, \textbf{regular}),
if there is a finite system $E$, (respectively, \textbf{regular} system),
of equations as above such that $T$ is the principal component of the least
solution of $E$ in $\tos(V_k)$.  
\end{definition}

An alternative definition is possible by interpreting finite systems of 
fixed point equations directly on the ``continuous categorical algebra'' 
of linear orders. See \cite{BEMezei}.  

We will use the above definition mainly when $k = 0$. It is known that 
when $T \in T^\omega_\Sigma(V_n)$ and $n < m$, then $T$ is algebraic in $T^\omega_\Sigma(V_n)$  
if and only if it is algebraic in $T_\Sigma(V_m)$. Thus, we may simply call $T$ 
just algebraic.  Moreover, we say that a tree 
$T \in T_\Sigma^\omega(V)$ is algebraic if it is algebraic 
in $T_\Sigma(V_n)$, for some $n\geq 0$. 
It is also known that when $\Sigma \subseteq \Sigma'$, then a tree $T\in T_\Sigma^\omega(V)$ is algebraic 
if and only if $T$ is algebraic in $T_{\Sigma'}^\omega(V)$. Similar facts and conventions hold for regular trees.
So below we can simply say that a tree is algebraic, or regular without specifying exactly the 
ranked alphabet.

The set of leaves of a tree $T$  in $\tos(V)$ is denoted
\begin{eqnarray*}
  \Fr(T) &=& \{u \in [n]^*: T(u) \in \Sigma_0 \cup V\} ,
\end{eqnarray*}
and is called the \textbf{frontier of $T$}. $\Fr(T)$ is a prefix language
linearly ordered by $\lex$.  Here, $n$ is the maximum of the ranks of the symbols in $\Sigma$.

\begin{definition}
A linear order is \textbf{algebraic} (respectively, \textbf{regular})  if it 
isomorphic to $(\Fr(T),\lex)$ for some algebraic (respectively, regular) tree. 
An algebraic or regular ordinal is an ordinal which is an algebraic or 
regular linear order. 
\end{definition}

Heilbrunner \cite{Heilbrunner} proved that the frontiers of regular trees
are those obtainable from the empty and one point  
frontiers by the operations of
``concatenation, omega and omega-op powers'', and infinitely many ``shuffle'' operations.
It is an easy corollary  
of this fact that the regular ordinals are those less than $\omega ^\omega $.
(There is a somewhat longer argument based only on the facts
in \cite{BlCho01}.) 
In this paper we will prove that an ordinal is algebraic if and only if
it is less than $\ooo$. 

The following fact is known. See \cite{CourcelleFund}.

\begin{proposition}
The classes of algebraic and regular trees are closed under first- and second-order substitution. 
\end{proposition}

In the remaining part of this section we show that 
for algebraic and regular linear orders 
one may restrict attention to those algebraic or regular $\Sigma $-trees for which 
$\Sigma_n=\emptyset$ unless $n=2$ or $n=0$.   

Let $\Delta$ be the ranked alphabet with one binary function symbol and
one constant symbol;
otherwise, $\Delta_n=\emptyset$.

\begin{proposition}
For any ranked alphabet $\Sigma $ and any algebraic tree $T\in T^\omega_\Sigma(V)$ there is an algebraic
tree $T'\in T^\omega_\Delta$ such that $(\Fr(T),\lex)$ and $(\Fr(T'),\lex)$ are isomorphic. 
\eop 
\end{proposition}

For example, consider the system $E$ of equations
\begin{eqnarray}
\nonumber
F_0 &=& \sigma_1(a,b,F_1(a))\\ 
\label{eq:example3}
F_1(x) &=& F_2(x,x)\\
\nonumber
F_2(x,y) &=& \sigma_1( \sigma_2(a),\ F_2(x,\ F_2(x,y)),\  y)
\end{eqnarray} 
which 
uses a function symbol $\sigma_1 $ in $\Sigma_3$.  The least
solution consists of three trees $(T_0,T_1,T_2)$ having vertices of out-degree 3.
We replace the system $E$ by the system
\begin{eqnarray*}
F_0 &=& g(a,\ g(a,F_1(a)))\\ 
F_1(x) &=& F_2(x,x)\\
F_2(x,y) &=& g (a,\ g( F_2(x,\ F_2(x,y)),\ y) )
\end{eqnarray*} 
in which the right hand terms 
use only the function variables and the one
binary function symbol $g $, and the one constant symbol $a$.
If $(T'_0, T'_1, T'_2)$ is the least solution of this second
system, 
$  (\Fr(T_i),\lex) $ is isomorphic to $(\Fr(T'_i), \lex)$, for $i=1,2,3$.

Thus, from now on, we will assume that if $T$ is an algebraic tree,
then $\Fr(T) \subseteq [2]^*$.

\begin{example}
\label{example-omega}
Let $\Sigma$ contain the binary symbol $g$, the unary symbol $f$ and the constant $a$.
Consider the system
\begin{eqnarray*}
F_0 &=& F(a)\\
F(x) &=& g (x, F(f(x)))
\end{eqnarray*} 
Then the first component of the least solution of this system is 
the tree 
$$T_0 = g(a,g(f(a),g(f(f(a)), \ldots ,g(f^n(a),\ldots )))).$$
Thus, this tree is algebraic. See also \cite{Courcelle,CourcelleFund,Guessarian}.  
\end{example}

\section{Closure properties of algebraic ordinals}
\label{sec-closure}

In this section we use certain closure properties of algebraic 
ordinals to prove that every ordinal less than $\ooo$ is 
algebraic.

\begin{proposition}\label{closure prop}
Let $\C$ be any set of ordinals which contains 0,1, and is
closed under sum, product and $\omega $-power: i.e., if
$\alpha , \beta \in \C$, then
$\alpha + \beta, \ \alpha \x \beta , \ \alpha^\omega $ belong
to $\C$. Then all ordinals less than $\ooo$ belong to $\C$.
\end{proposition}

 {\sl Proof.\/ } 
This follows from the assumptions and induction, making use
of the Cantor Normal Form \cite{Roitman,Rosenstein,Sierpinski} 
for ordinals less than $\ooo$. 
In fact, it is known that the set of ordinals less than $\ooo$ is the least
set of ordinals containing $0,1$ which is closed under sum, product, and $\omega$-power. 
\eop

We use Proposition~\ref{closure prop}  to show all ordinals less than $\ooo$ are
algebraic.  For this reason, we fix the ranked alphabet 
$\Delta $ containing only a binary symbol $g$ and a unary symbol $a$. 
We show that all ordinals less than $\ooo$ 
arise as frontiers of algebraic trees in $T_\Delta^\omega$.

The ordinal 0 is algebraic, since if $T$ is the empty tree, then
$\Fr(T)$ is the empty language, and 
the order type of the empty language is 0.
The ordinal 1 is algebraic since the one-point tree is algebraic. 

Suppose that $\alpha=\o(\Fr(T),\lex)$ and $\beta=\o(\Fr(S),\lex)$,
where $T$ and $S$ are algebraic trees in $T_\Delta^\omega$. 
\begin{proposition}
\label{sum prop}
If $\alpha $ and $\beta $ are algebraic ordinals, so is $\alpha + \beta $.
\end{proposition}
{\sl Proof.\/ } 
Consider the algebraic tree $g(T,S) = g(a,b)[a \mapsto T,b \mapsto S]$
whose root is labeled by the function symbol $g$
and whose left subtree is $T$ and whose right subtree is $S$. 
Then the tree $g(T,S)$ 
is algebraic, and its frontier has order type $\alpha + \beta$. \eop   

Since $0$ and $1$ are algebraic, 
we have the easy corollary
that \textit{every finite ordinal is algebraic}.

\begin{proposition}
\label{product prop}
If $\alpha $ and $\beta $ are algebraic ordinals, so is $\alpha \x \beta $.
\end{proposition}
{\sl Proof.\/ } 
The tree $S[a \mapsto T]$ is
algebraic, and its frontier has order type $\alpha \times \beta$. 
\eop

\begin{proposition}
\label{omega pow prop}
  If $\alpha $ is an algebraic ordinal, so is $\alpha ^\omega $.
\end{proposition}
{\sl Proof.\/ } Suppose that $\alpha = \o(\Fr(T),\lex)$ for 
an algebraic tree $T$ in $T_\Delta^\omega$. 
Consider the tree $T_0$ of Example~\ref{example-omega} and let $S$ 
be the algebraic tree in $T_\Delta^\omega$ obtained
by substituting the tree $T$ for each vertex labeled $f$:
$S = T_0[f \mapsto T]$. 
Then $S$ is algebraic and its frontier is of order type $\alpha^\omega$. 
\eop

\begin{corollary}
 Every ordinal less than $\ooo$ is algebraic.
\end{corollary}
{\sl Proof.\/ } 
By Propositions \ref{sum prop}, \ref{product prop} and \ref{omega pow prop}, together
with Proposition \ref{closure prop}. \eop

\section{Grammars}\label{sec-grammars} 

In our argument proving that all algebraic ordinals are less than
$\ooo$ we will use certain context-free grammars, called 
ordinal grammars.

Throughout this section, we assume that
$$G=(N,T,S,P)$$ 
is a context-free grammar,
with nonterminals $N$, start symbol $S \in N$, terminals
$T=\{0,1\} $, and productions $P$.  

We will denote finite words  
on the alphabet $\{0,1\}$ by $u,v$, $w,x,y,\ldots$;
nonterminals will be written $X,Y,Z, \ldots$, and 
we will denote by $p,q,r,s \ldots $
words on $N \cup T$, 
possibly containing nonterminals.

Further, we assume that each context-free grammar has
the following properties:
\begin{itemize}

\item Either each nonterminal $X \in N$ is ``coaccessible'',
i.e., $L(X)$ is
a nonempty subset of $\zos$, where 
$L(X)$ is the collection of all words $u \in \zos$
such that there is some derivation
\begin{eqnarray*}
X &\dv & u,
\end{eqnarray*}
or $N = \{S\}$ and $P$ is empty.
We write $L(G)$ for $L(S)$.
\item  Each nonterminal $X$ is ``accessible'', i.e., there is
some derivation 
\begin{eqnarray*}
 S &\dv & { q   X r  }
\end{eqnarray*}
where $ q ,\ r \in (N \cup T)^*$.
\end{itemize}

We end this section with some classical definitions.
Suppose $X,Y$ are nonterminals. 
Write
\begin{eqnarray*}
 Y &\preceq & X 
\end{eqnarray*}
if there is some derivation $X \dv pYq $ 
for some $p $ and $ q $. Define $X \approx Y$ if both 
$X \preceq Y$ and $Y \preceq X$ hold. When $Y \preceq X$ but $X \approx Y$ 
does not hold, we write $Y \prec X$. 

The relation $\preceq$ is a preorder on the nonterminals, and
induces a partial order $\leq$ 
on the equivalence classes 
\begin{eqnarray*}
  [X] &:=& \{Y: X \approx Y\} ,
\end{eqnarray*}
where $[Y]\leq [X]$ if $Y \preceq X$.

We say \textbf{$X$ is a recursive nonterminal} 
if there is some \textit{nontrivial derivation}  
\begin{eqnarray}
\label{circular}
  X & \dv & p X q  ,
\end{eqnarray}
for some words  $p,q \in (N \cup T)^*$.
If not, \textbf{$X$ is a non-recursive nonterminal}.
When $X$ is non-recursive, and $X \to p  $ is a rule,
$Y \prec X$, for all nonterminals in $p   $.  

\begin{definition} 
The \textbf{height} of a nonterminal $X$ is the number
of equivalence classes
 $[Y]$ strictly below $[X]$.    
If $ q  $ is a finite 
word on $N \cup T$, the height of $ q  $,
$\hgt( q  )$, is the maximum of the heights of
the nonterminals occurring in $ q  $.  If $ q  $
has no nonterminals, $\hgt( q  )= -1$.
\end{definition}
If there are $k$ 
nonterminals, $\hgt(X) < k$, for all nonterminals $X$. 
If $X$ has height zero,
and $Y \preceq X$, then $Y \approx X$.

For any word $ q  $ in $(N \cup T)^*$,
write $L( q )$ for all words in $\zos$ derivable in $G$ from
$ q  $.

\subsection{Prefix and ordinal grammars}

Our definition of an ordinal grammar is motivated by 
Proposition~\ref{prefix prop} below. But first we need
the following fact:  

\begin{lemma}
\label{strict lem}
Suppose $L \subseteq \zos$ is any language. If $(L,\lex)$ is not
well-ordered, then there is a countable descending chain $(u_n)$, $n=0,1,\ldots $,
of words in $L$ such that
\begin{eqnarray*}
  u_{n+1} & <_s & u_n,
\end{eqnarray*}
for each $n \geq 0$.
\end{lemma}
{\sl Proof.\/ } Suppose that $(v_n)$ is a countable $\lex$-descending
chain of words in $L$.  Then, for each $n$, either $v_{n+1} <_p v_n$
or $v_{n+1} <_s v_n$.  Now define $u_0=v_0$. Since $v_0$ has only
finitely many prefixes, there is a least $k$ such that $v_{k+1} <_s v_k <_p \ldots <_p v_0$.  
Then $u_1=v_{k+1} <_s u_0$, since $u<_s v$ if $u<_s w$ and $w <_p v$, for
any words $u,v,w$.
Similarly, assuming that $u_m$ has  
been defined as $v_{m'}$, for some $m'$, we may define $u_{m+1}$ as the first
$v_k$ such that $k>m'$ and $v_k <_s u_m$.  \eop 

We take note of the following inheritance property.
\begin{proposition}
\label{prefix prop}
If $(L(G),\lex)$ is well-ordered, then, for any nonterminal $X$,
$(L(X), \lex)$ is also well-ordered.
\end{proposition}
{\sl Proof.\/ } If not, by Lemma
\ref{strict lem}, 
suppose that there is a countable chain 
\begin{eqnarray*}
  \ldots <_s u_1 <_s u_0.
\end{eqnarray*}
of words in $L(X)$. Thus, for each $n$, 
$u_{n+1} = x_n0y_n$ and $u_n=x_n1z_n$, for some words $x_n,y_n,z_n \in \zos$.
Since all nonterminals are accessible and coaccessible,
there are words $v,w$ such that
\begin{eqnarray*}
  v u_i w & \in & L(G),
\end{eqnarray*}
for each $i \geq 0$, and thus
\begin{eqnarray*}
  v u_{n+1} w &<_s& v u_{n} w,
\end{eqnarray*}
for each $n \geq 0$, showing $(L(G),\lex)$ is not
well-ordered, contradicting the hypothesis. \eop

\begin{definition}
  \label{def: ordinal grammar}
A grammar $G$ is a \textbf{prefix grammar}
if, for \textit{each} nonterminal $X$, $L(X)$ is a prefix language.  
An \textbf{ordinal grammar} is a prefix grammar such that
$(L(G), \lex)$ is well-ordered.
\end{definition}
If $G = (N,T,P,S)$ is an ordinal grammar, then by Proposition~\ref{prefix prop}, 
 $(L(X),<_\ell) = (L(X), <_s)$ is well-ordered, for all $X \in N$.

The following fact is immediate from Lemma~\ref{pref lemma} and the definitions.
\begin{lemma}
\label{lem-ord grammar}
If $G$ is a prefix grammar, then for any word $ q = v_0X_1v_1\ldots v_{k-1}X_kv_k \in (N \cup T)^*$, 
$L(q)$ is a prefix language with $(L( q ), \lex) = (L(q), <_s)$ and  
$$\o(L(q),\lex) = \o(L(X_k),\lex)\times \ldots \times  \o(L(X_1),\lex).$$
Moreover, if  $G$ is an ordinal grammar,
then for any word $ q  \in (N \cup T)^*$, $(L( q ), \lex)$  is well-ordered.
\end{lemma}

We write $\o( q )$ for the order type of $(L(q), \lex)$. In particular,
$\o(X)$ denotes the order type of the linear order $(L(X), \lex) = (L(X),<_s)$.

We list some examples of ordinal grammars.  Each grammar
includes  nonterminals from the set
$$\{\Omega_1, \Omega_2,\ldots, \Omega^\Omega_1, \Omega^\Omega_2, \ldots \} $$
 and productions of some of the previous ones.
\begin{enumerate}
\item $\omega $
  \begin{eqnarray*}
\Omega_1 & \to & 0 + 1 \Omega_1 .
  \end{eqnarray*}
$L(\Omega_1)=1^*0$, so that $\o(\Omega_1) = \omega $.
\item $\omega ^2$
  \begin{eqnarray*}
\Omega_2 & \to & 0 \Omega_1  + 1 \Omega_2 .
  \end{eqnarray*}
$L(\Omega_2)= \bigcup_{k,n} 1^k 0 1^n 0 = L(\Omega)^2$, so
that $\o(\Omega_2)=\omega^2$.
\item $\omega^{n+1}$
  \begin{eqnarray*}
\Omega_{n+1} & \to & 0 \Omega_n  + 1 \Omega_{n+1}.
  \end{eqnarray*}
$\o(\Omega_{n+1})=\o(\Omega_1)^{n+1}$.
\item $\omega^\omega $
  \begin{eqnarray*}
\Omega_1^\Omega  & \to & 0 + 1 \Omega_1 ^\Omega \Omega_1 .
  \end{eqnarray*}
$L(\Omega_1^\Omega)=\bigcup_n 1^n 0 L(\Omega_1)^n = \bigcup_n 1^n 0 (1^* 0)^n$, 
so that
$\o(\Omega_1^\Omega)=\omega^\omega $.
\item $\omega^{\omega^{n+1}}$
  \begin{eqnarray*}
\Omega_{n+1}^\Omega  & \to & 0+ 1 \Omega^{\Omega}_{n+1} \Omega_n^\Omega 
  \end{eqnarray*}
\end{enumerate}
Note that the first three  are regular grammars, 
but the subsequent ones are context-free ordinal grammars.

We now establish some closure properties of the ordinals of ordinal grammars. 

\begin{proposition}
\label{ord grammar closure prop}
The set of ordinals $\o(L(G))$, for an ordinal grammar $G$,
is closed under sum, products and  $\omega $-powers. 
\end{proposition}

{\sl Proof.\/ } Suppose that $G_i$ are ordinal grammars, for $i=1,2$.
Then, the grammar with a new start symbol $S_{+}$ and productions
\begin{eqnarray*}\label{}
S_{+} & \to & 0S_1 + 1 S_2
\end{eqnarray*} 
together with the productions of $G_1$ and $G_2$ is a grammar satisfying
\begin{eqnarray*}\label{}
\o(S_{+}) &=& \o(L(G_1)) + \o(L(G_2)). 
\end{eqnarray*} 
If, instead, we add the new start $S_{\x}$ and the production
\begin{eqnarray*}\label{}
S_{\x} & \to & S_2 S_1
\end{eqnarray*} 
to the productions of $G_1$ and $G_2$, we obtain  a grammar satisfying
\begin{eqnarray*}\label{}
\o(S_{\x}) &=& \o(L(G_1)) \x \o(L(G_2)).  
\end{eqnarray*}
Last, if we add the new start symbol $S_\omega $ and the productions
\begin{eqnarray*}\label{}
S_{\omega } & \to & 0 + 1 S_\omega S_1
\end{eqnarray*} 
to the productions of $G_1$, we obtain a grammar satisfying
\begin{eqnarray*}\label{}
\o(S_{\omega }) &=& \o(L(G_1))^\omega .  
\end{eqnarray*} 
In each case, the constructed grammar is an ordinal grammar.
\eop

\begin{proposition}
Any ordinal less than $\ooo$ is
the ordinal  $\o(L(G))$, for some ordinal grammar $G$.
\end{proposition}
{\sl Proof.\/ } From Proposition \ref{ord grammar closure prop} and Proposition
\ref{closure prop}, using the fact that every finite 
ordinal is the ordinal of an ordinal grammar.  \eop

\begin{remark}
The above proposition also follows from the corresponding result for algebraic ordinals and 
the facts in Section~\ref{sec-translation}. 
\end{remark}

\subsection{A bound on ordinals of ordinal grammars}  
In this section, we will show that
for any ordinal grammar $G$, $\o(L(G))$ is less than $\omega^{\omega^\omega}$. 

\textit{Throughout this section, we assume that $G$ is an ordinal grammar and $L(G)$ is infinite.
Moreover, we assume that $L(X)$ 
contains at least two words, for each nonterminal $X$.} 
Note that for each nonterminal $X$, since
$L(X)$ is a prefix language, it does not contain the empty word.

\begin{proposition}
\label{ord prop1}
If $X \dv q$ for some $X \in N$ and  $q =  v_0 X_1 v_1 \ldots v_{k-1} X_k v_{k} \in (N \cup T)^*$
then $\o(q) = \o(X_k)\times \ldots \times \o(X_1) \leq \o(X)$. In particular, 
$\o(X_i) \leq \o(X)$ for all $i$. 
\end{proposition}
{\sl Proof.\/ } Immediate from Lemma~\ref{lem-ord grammar} and the fact that $L(q) \subseteq L(X)$. \eop

The relations $\preceq$ and $\approx$ were defined at the beginning of Section~\ref{sec-grammars}.

\begin{corollary}
\label{dominates cor}
If $ Y \preceq X$, then $\o(Y) \leq \o(X)$. If $X \approx Y$, $\o(X)=\o(Y)$.
\end{corollary} 
{\sl Proof.\/ } Immediate, from Proposition \ref{ord prop1}. \eop

It is not necessarily true that if $\o(X)=\o(Y)$ then $X \approx Y$.  
Consider the example:
\begin{eqnarray*}
  X_1 & \to & 1 X_2  \\
  X_2 & \to & 1 X_2 + 0.
\end{eqnarray*}
Then $X_2\prec X_1$, and
$\o(X_1)=\o(X_2)=\omega $, since
\begin{eqnarray*}
  L(X_2) &=& \bigcup_{n \geq 0} 1^n 0 \\
L(X_1) &=& \bigcup _{n \geq  1} 1^n 0 .
\end{eqnarray*}

\begin{proposition}
\label{no left recursive}
For any nonterminal $X$, 
there is no derivation $X \dv Xp$ with $p \neq \epsilon $. 
\end{proposition}

{\sl Proof.\/ } 
Indeed, suppose that $X \dv X p  $. There are
nonempty words $u,v$ such that  $p  \dv v$, and $X \dv u$.
Thus, $X \dv u$ and $X \dv uv$, so that
$L(X)$ is not a prefix language. \eop

\begin{proposition} 
\label{first occ prop}
For each derivation $X \dv p  $, 
either $Y \prec X$ holds for all nonterminals $Y$ in $p $, 
or there is exactly one nonterminal $Y$ occurring in $p $ with 
$X \approx Y$, and in this case,
$p  =uY q $, for some $u \in \zos$ and for some $q$ with $\hgt( q )<\hgt(X)$.
\end{proposition}
{\sl Proof.\/ } 
Suppose that $X \dv  q Y r Z s$ is a derivation, where $Y,Z \approx X$.
Let $\alpha=\o(X),\ \beta = \o(Y),\ \gamma=\o(Z)$. Then by Corollary~\ref{dominates cor},
$\alpha = \beta = \gamma$, and thus by Proposition~\ref{ord prop1}, 
$\alpha \times \alpha \leq \o(q Y r Z s) \leq \alpha$. 
This contradicts the assumption that $L(X)$ contains at least two words.
Similarly, if there are nonterminals $Y,Z$ such that $X \dv  q Y r Z s$   
with $\hgt(Y) < \hgt(X)=\hgt(Z)$, then $\o(X) \geq  \o(X) \x \o(Z) \geq \o(X) \x 2 $, 
an impossibility, since $\o(X) \geq 2$. 
\eop

The next fact is a basic result.
\begin{lemma}
\label{incomp lem}
Suppose  $X \dv u X p  $ and $X \dv v X  q  $.
If  $|u| \leq |v|$, then $u \leq_p v$.
In particular, if $|u|=|v|$, then
$u=v$.
\end{lemma} 
{\sl Proof.} Assume $u <_s v$, say. 
Suppose $p \dv w$ and $ q  \dv w'$. Then, 
for each $n$, $X \dv u^n\, X\, w^n$, so that
$X \dv u^n\, v\, z\, w'\, w^n = y_n$, where $z$ 
is any word in $\{0,1\}^*$ with $X \dv  z$. 
But, $y_{n+1}<_s y_n$, for each $n$, contradicting
the fact that $(L(X),\lex)$ is well-ordered. \eop

\begin{lemma}
 \label{hierarchy lemma}
Suppose there is a derivation $X \dv u X p$. Suppose also that
$X \dv v$. Then either $v <_s u$ or $u <_p v$.
\end{lemma}
{\sl Proof.\/ } 
For any two words $u,v$ there are four possibilities:
$$u <_s v, \ v \leq_p u,\ v<_su, \ u <_p v.$$
We show that the first two possibilities
are ruled out.

If $u <_s v$, then, for any $n \geq 1$,
 $ u^{n+1} v w^{n+1} <_s u^n v w^n$, where $w$ is any terminal word 
with $p \dv w$, so that there is a descending
chain in $L(X)$.

If $v \leq_p u$, then since $v \neq \epsilon$, $v <_p uvw$, for any word $w$, so that $L(X)$ is
not a prefix language. \eop

Recall that a \textit{primitive word} is a nonempty word $v$ which cannot
be written as $u^n$, for any word $u$ and integer $n>1$.  
The \textbf{primitive root} \cite{Lothaire} of a nonempty word $v$ is a primitive word
$u$ such that $v=u^n$, for some $n \geq 1$.
\begin{proposition}
\label{u0 prop}
Suppose that $X$ is a recursive nonterminal. Then there is a 
unique shortest word $u_0^X\in \zop $ such that whenever $X \dv  uXp$
for some $u \in \zop $ and $p\in (N \cup T)^*$, then $u$ is a power 
of $u_0^X$.
\end{proposition}
{\sl Proof.} Consider any derivation $X \dv vXq$ with 
$v \in \zop $ and $q \in (N \cup T)^*$, let $u_0^X$ denote the primitive 
root of $v$. Thus, $u_0^X$ is the shortest word such that $v$ is a 
power of $u_0^X$, and clearly, $u_0^X$ is primitive. 
If $X \dv  uXp$ then there are some $m,n\geq 1$
such that $|u^m|=|v^n|$. But then, by Lemma \ref{incomp lem},
\begin{eqnarray*}
  u^m &=& v^n.
\end{eqnarray*}
It then follows that $u,v$ are powers of the same word
(see \cite{Lothaire} for example), 
which implies that $u$ also is a power of $u_0^X$. \eop 
   
Below we will write $u_0$ for $u_0^X$ whenever $X$ is clear from the context.
\begin{proposition}
Suppose that $X$ is recursive and $v \in L(X)$. Then, there
is some $n \geq 1$ such that  $v <_s u_0^n$.
\end{proposition} 
{\sl Proof.\/ } Indeed, there is a word
$u$ such that $|v|<|u|$ and $X \dv u X p$, for some $p \in (N \cup T)^*$.
Then, $v <_s u$, by Lemma \ref{hierarchy lemma},
and $u = u_0^n$, for some $n$, by Proposition \ref{u0 prop}. \eop 

Thus, for $v \in L(X)$, if we choose
$n$ as the least integer such that $v <_s u_0^n$, we
may write $v$ in a unique way as $u_0^{n-1} w$ 
where $u_0$ is not a prefix of $w$ and $w <_s u_0$. Moreover, we 
can write $w$ as $x0y$ where $x1$ is a prefix of $u_0$.

\begin{definition}
Suppose that $X$ is 
a recursive nonterminal, and the word 
$x1$ is a prefix of $u_0$. For each $n \geq 0$, 
define $L(n,x,X)$ as the set of all words of the form $u_0^nx0y$
in $L(X)$. Moreover, define $L(n,X) = \bigcup_x L(n,x,X)$, where $x$ 
ranges over all words such that $x1$ is a prefix of $u_0$.
\end{definition} 

\begin{lemma}
\label{sum lemma}
Suppose $n<m$.  If $v \in L(n,X)$ and $w \in L(m, X)$, then
$v  <_s w$. 
\end{lemma}
{\sl Proof.\/ } 
Write $v=u_0^n x0y$ and $w=u_0^{m}x'0y'$, where $x1$ and $x'1$ are
prefixes of $u_0$.
If $u_0=x1r$, then
\begin{eqnarray*}\label{}
v &=&  u_0^n x0 y \\
 &<_s&  u_0^{n} (x1r)^{m-n} x'0y' \\
&=& u_0^{m} x'0y' \ = \ w. \eop 
\end{eqnarray*} 

The following lemma gives an easy upper bound to the ordinal of a union.

\begin{lemma}\label{upper bound lemma}
Suppose that  $L_1, L_2$ are subsets of $\zos$ such that
for $i=1,2$, $(L_i,\lex)$ is well-ordered. 
Then $(L_1 \cup L_2, \lex)$ is well-ordered. 
Let $\o(L_i,\lex) \leq \alpha_i$, where $\alpha_i$ is a (infinite) limit ordinal, 
so that $1 + \alpha_i=\alpha_i$,
$i = 1,2$. Then
\begin{eqnarray*}
\o(L_1 \cup L_2, \lex) & \leq & \max\{\alpha_1 \x \alpha_2, \alpha_2 \x \alpha_1 \}. 
\end{eqnarray*} 
\end{lemma} 
{\sl Proof.\/ } To show $L_1 \cup L_2$ is well-ordered, suppose that
$(v_n)$ is an infinite descending chain in $L_1 \cup L_2$.  Then,
either there are infinitely many $v_n \in L_1$, or infinitely many $v_n$
in $L_2$.  Either possibility contradicts the assumption that
both $L_1$ and $L_2$ are well-ordered.

Without loss of generality we may assume that $(L_1,\lex)$ is 
cofinal in $(L_1 \cup L_2,\lex)$, i.e., for each $y \in L_2$ 
there is some $x \in L_1$ with $y \leq x$. For each $y \in L_2$ 
let $f(y)$ denote the least $x \in L_1$ with $y \leq x$.
For each $x \in L_1$, let $\beta(x)$ denote the 
order type of the set $\{y \in L_2 : f(y) = x\}$. 
Then, using Lemma~\ref{lem-ordinals} in the last line, 
\begin{eqnarray*}
\o(L_1 \cup L_2)
&\leq& \sum_{x \in L_1} (\beta(x) + 1)\\
&\leq& \sum_{x \in L_1} (\alpha_2 + 1) \\
&= & (\alpha_2 +1 )\times \alpha_1 \\
&\leq& \alpha_2 \x \alpha_1.
\end{eqnarray*}
Thus, in this case, $\o(L_1 \cup L_2) \leq \alpha_2 \x \alpha_1$. \eop

By induction, we have:
\begin{corollary}
\label{upper bound cor}
For any finite collection $\{L_i: i =1,\ldots,n\}$  
of subsets of $\zos$ such that 
for each $i$, $(L_i,\lex)$ is well-ordered, $(\bigcup_i L_i, \lex)$ is well-ordered.
Moreover, if $\o(L_i,\lex) \leq \alpha_i$ where $\alpha_i$  is a limit ordinal
for all $1 \leq i \leq n$,  it holds that  
$(\bigcup_i L_i, \lex)$ 
\begin{eqnarray*}
  \o( \bigcup_i L_i, \lex) & \leq & \max_\pi \{ \alpha_{\pi(1)} \x \ldots \x \alpha_{\pi(n)}\}
\end{eqnarray*}
where $\pi $ ranges over all permutations of $\{1,\ldots,n\}$. \eop 
\end{corollary}

The next theorem is one of our main results. 
\begin{theorem}
\label{height thm}
Suppose that $X$ is a nonterminal of height $h$. Then 
\begin{eqnarray*}
 \o(X) &\leq & \omega^{\omega^h} .
\end{eqnarray*}
\end{theorem}
{\sl Proof.\/ } 
We prove this claim by induction on the height of $X$.
Let $X$ be a nonterminal of height $h$ and suppose that
we have proved the claim for all nonterminals of height less than $h$. 
Below we will make use of the fact that the set of ordinals less than  
$\omega^{\omega^h}$ is closed under sum and product. Moreover,
when $h > 0$, then for every ordinal $\alpha < \omega^{\omega^h}$ 
there is a limit ordinal $\beta < \omega^{\omega^h}$ with $\alpha < \beta$.
Indeed, we can choose $\beta = \omega^{h-1}\times n$ for some $n$.

Case 1. $X$ is not recursive. Then, by the induction hypothesis, we 
have $\o(Y) < \omega^{\omega^h}$ whenever $Y$ occurs on the 
right side of a production whose left side is $X$. It follows 
by Proposition~\ref{ord prop1} 
that $\o(p)  < \omega^{\omega^h}$ whenever $X \to p$ is a production.
Since $L(X)$ is a finite union of the languages $L(p)$, it follows 
that $\o(X)  < \omega^{\omega^h}$, by Corollary \ref{upper bound cor}.
(If $h=0$, $L(p)$ is a single word in $\zos$. Moreover, $\o(X)$ is finite.)

Case 2. $X$ is recursive. Then for each $n,x$, $L(n,x,X)$ 
is a finite union of languages of the form $L(u_0^nx0p)$, where 
there is left derivation $X \dv u_0^n x 0 p$. 
It is not possible that $p$ contains a nonterminal $Y$ with
$X \approx Y$, since it that case we would have a derivation
$X \dv u_0^nx0wXq$ for some terminal word $w$ and 
some $q$, contradicting Proposition~\ref{u0 prop}. 
Thus, $Y \prec X$ holds for all nonterminals $Y$ 
occurring in $p$. It follows by the induction hypothesis that $\o( L(u_0^nx0p),<_\ell )
< \omega^{\omega^h} $. Thus, using Corollary \ref{upper bound cor},
it follows that $\o( L(n,x,X),<_\ell ) < \omega^{ \omega^h }$. 
Again, by Corollary \ref{upper bound cor},
\begin{eqnarray*}
 \alpha_n &=&  \o( L(n,X),<_\ell) \\
  & < & \omega^{\omega^h}.
\end{eqnarray*}
(If $h=0$, $\alpha_n < \omega $, for all $n$.)
But, by Lemma \ref{sum lemma}, 
\begin{eqnarray*}
  \o(X) &=& \alpha_0 + \alpha_1 + \ldots \\
&=& \sup \{\sum_{i \in [n]} \alpha_i : n \geq 0\}.
\end{eqnarray*}
Since $\sum_{i \in [n]} \alpha_i < \omega^{\omega^h}$ for all $n$, 
it follows that $\o(X)\leq \omega^{\omega^h}$.  \eop

\begin{corollary}
  If $G$ is an ordinal grammar, there is an integer $n$ such that 
  \begin{eqnarray*}
    \o(G) & \leq & \omega ^{\omega ^n}.
  \end{eqnarray*}
\end{corollary}
{\sl Proof.\/ } Let $n$ be the height of the start symbol $S$.
Then, by Theorem
\ref{height thm}, $\o(S) \leq \omega^{\omega ^n}$. \eop 

We have thus completed the proof of this characterization of the ordinals of
ordinal grammars:
\begin{theorem}
\label{main thm}
 An ordinal $\alpha $ is less than $\ooo$ if and only if
there is some ordinal grammar $G$ such that
$\alpha = \o(L(G),\lex)$.
\end{theorem}

\section{From algebraic trees to prefix grammars}\label{sec-translation}

In this section we show that each system of equations defining an
algebraic tree can be transformed (in polynomial time) to a prefix
grammar\footnote{Here, we allow grammars over an 
arbitrary (linearly ordered) terminal alphabet. The notion of a prefix grammar 
can be adjusted appropriately.}  generating the frontier of the tree. This result allows 
us to complete the proof of the fact that every algebraic
ordinal is less than $\ooo$.

Consider a system of equations
\begin{eqnarray}
\label{a system}
F_i(x_0,\ldots,x_{n_i-1}) &=& t_i(x_0,\ldots,x_{n_i-1}), \quad i=1,\ldots,m,   
\end{eqnarray}
where each $t_i$ is a term over the ranked alphabet $\Sigma \cup \F$
in the variables $x_0,\ldots,x_{n_i-1}$. 
We assume that $F_1$ is the principal function variable and that $n_1 = 0$. 
Each component of the least  solution is an algebraic tree. 
Let $(T_1, T_2,\ldots, T_m)$ denote the least solution of the 
system (\ref{a system}). 

For the ranked alphabet $\Sigma \cup \F$, let $\Delta$ be the 
(unranked) alphabet whose letters are the 
letters $(\sigma,k), (F_i,j)$ 
where $\sigma \in \Sigma_n$, $n > 0$ and $k \in [n]$, 
$j \in [n_i]$ and $1 \leq i \leq m$. Let 
$T$ be a finite or infinite tree in $T_{ \Sigma \cup \F}^\omega(V)$.

\begin{definition}
\label{hat definition}
For each vertex $u \in \dom(T)$ we 
define a word $\widehat{u} \in \Delta^*$ by induction.
First, $\widehat{\epsilon } = \epsilon$.  When $u = vi$ and $T(v) = \delta $, then 
$\delta \in ( \Sigma \cup \F)_k$, for some $k>0$, and we define
$\widehat{u} = \widehat{v}(\delta,i)$.   

We define the ``labeled frontier language of $T \in T_\Sigma^\omega$'' as
the set of words
$$\Lfr(T) = \{\widehat{u}T(u) : T(u) \in \Sigma _0\}$$ 
\end{definition}
We prove the following.
\begin{theorem}
\label{alg to prefix}
When $T \in T_\Sigma^\omega$ is an algebraic tree, $\Lfr(T)$ can be generated by a 
prefix grammar.\footnote{In \cite{CourcelleFund}, $\Lfr(T)$ is called the \emph{branch 
language} of $T$.  Courcelle showed
that a ``locally finite'' tree $T$ is algebraic if and only if 
 $\Lfr(T)$ is a strict deterministic context-free language, see \cite{Courcelle,CourcelleFund}.} 
\end{theorem}
{\sl Proof.\/ } 
Suppose that $T$ is the principal component of the least solution of the 
system (\ref{a system}).

We will define a grammar whose nonterminals $N$ consist of the 
letters $F_i$, together with all ordered pairs $(F_i,j)$ where 
$i = 1,\ldots,m$, $j \in [n_i]$.

The grammar will be designed to have the following properties.

\textbf{Claim:}
\textit{For any word $u$, $T_i(u) = x_j$ if and only if $(F_i,j) \dv \widehat{u}$.
And for any word $u$, $T_i(u) \in \Sigma_0$ if and only if $F_i \dv \widehat{u}T_i(u)$.
Moreover, any terminal word derivable from $(F_i,j)$ is of the form $\widehat{u}$, 
and any terminal word derivable from $F_i$ is of the form $\widehat{u}T_i(u)$ 
for some $u \in \dom(T_i)$.} 

Let $\Gamma = \Sigma_0 \cup \{(\sigma,j): \sigma \in \Sigma_k,\ j \in [k]\}$.
The grammar generating $\Lfr(T)$ is: $G_L= (N,\Gamma,P,F_1)$, 
where $N = \F \cup \{(F_i,j) : 1 \leq i \leq m,j\in [n_i]\}$ and 
the set $P$ of productions is defined below. 
If $t_1,\ldots, t_m$ are the terms on the right side of (\ref{a system}) above, 
then the productions are: 
\begin{itemize} 
\item 
$$(F_i,j) \to \widehat{u}$$
where $u \in \dom(t_i)$ and $t_i(u) = x_j$,
\item 
$$F_i \to \widehat{u}t_i(u)$$
where  $u \in \dom(t_i)$ and  $t_i(u) \in \Sigma_0 \cup \F$.
\end{itemize}
The proof of the fact that the above grammar is a prefix grammar 
generating the language $\Lfr(T)$ relies on the above claim,
and may be found in the Appendix.
\eop 

\textbf{Example.} Suppose the system of equations is: 
\begin{eqnarray*}
F_0 &=& F_1(a)\\
F_1(x) &=& F_2(a,x)\\
F_2(x,y) &=& \sigma(x,a,F_2(x,F_2(x,y)))
\end{eqnarray*} 
where the individual variables $x,y$ stand for $x_0, \ x_1$, respectively.
Then the productions in $G_L$ are: 
\begin{eqnarray*}
F_0 &\to& F_1 + (F_1,0)a\\
F_1 &\to& F_2 + (F_2,0)a\\
(F_1,0) & \to & (F_2,1)  \\ 
(F_2,0) &\to&  (\sigma,0) + (\sigma,2)(F_2,0) + (\sigma,2)(F_2,1)(F_2,0) \\
(F_2,1) &\to& (\sigma,2)(F_2,1)(F_2,1) \\ 
F_2 &\to&   (\sigma,1)a + (\sigma,2)F_2 + (\sigma,2)(F_2,1) F_2\\
\end{eqnarray*}

\begin{corollary}
For every system of equations defining an algebraic tree $T \in T_\Sigma^\omega$ 
one can construct in polynomial time a 
prefix grammar generating the frontier of $T$. 
\end{corollary} 
{\sl Proof.\/ }
To get from the prefix grammar $G_L$ which derives $\Lfr(T)$ to 
a grammar $G'$ which derives $\Fr(T)$, replace each letter $(\sigma,j)$ by
just $j$, and delete the constant symbols in $\Sigma_0$.  Thus, in
the example above, the productions
of $G'$ are
\begin{eqnarray*}
F_0 &\to& F_1 + (F_1,0)\\
F_1 &\to& F_2 + (F_2,0)\\ 
(F_1,0) & \to &  (F_2,1) \\
(F_2,0) &\to& 0 + 2 (F_2,0) + 2(F_2,1)(F_2,0)\\
(F_2,1) &\to& 2(F_2,1)(F_2,1)\\ 
F_2 &\to&  1 + 2F_2 + 2(F_2,1)F_2.
\end{eqnarray*}
It follows that $G'$ is a prefix grammar generating
$\Fr(T)$.  \eop

\begin{corollary}
\label{alg to ordinal grammar}
If $\alpha $ is an algebraic ordinal, there is an ordinal grammar
$G'$ with $\alpha = \o(L(G'))$. 
\end{corollary}

We may now derive our main theorem.
\begin{theorem}
  An ordinal is algebraic if and only if it is less than $\ooo$.
\end{theorem}

We needed only to prove the ``only if'' direction.  
But
this follows immediately from Corollary
\ref{alg to ordinal grammar} and Theorem
\ref{main thm}. \eop

\section{Conclusion}
\label{sec-conclude}

We have proved that the algebraic ordinals are exactly those less than
$\omega^{\omega^\omega}$, or equivalently, the ordinals that can 
be constructed from $0$ and $1$ by the sum and product operations, and the 
operation $\alpha \mapsto \alpha^\omega$. It is  known that the  
regular ordinals  are those less than ${\omega^\omega}$, or equivalently,
those that can be constructed from $0,1$ and $\omega$ by just  sum and
product; or the ordinals that can be constructed from $0$ and $1$ by sum, 
product, and the operation $\alpha \mapsto \alpha \times \omega$. 

Recall (from \cite{Rosenstein}, for example) 
that the \textbf{Hausdorff rank} of a countable \emph{scattered} 
linear ordering  $L$
is the least ordinal $\alpha $ such that $L \in V_\alpha $, where
\begin{eqnarray*}\label{}
V_0 &:=& \{0, 1\}, 
\end{eqnarray*}
and if $\alpha>0$, the collection $V_\alpha $ is defined by:
\begin{eqnarray*}\label{}
V_\alpha &=& \{\sum_{i \in I} L_i: \ L_i \in \bigcup_{\beta<\alpha} V_\beta  \} ,
\end{eqnarray*}
where $I$ is either $\omega $, or $ \omega^*$, the reverse
of $\omega $, or a finite ordinal $n$, or $ \omega^*+\omega $.

In order to characterize the algebraic \textit{linear orders}, 
the next step might be the characterization of algebraic \textit{scattered} linear orders.  
We conjecture that any such linear order has 
Hausdorff rank less than $ \omega^\omega$. Moreover,
one possible conjecture is that, up to isomorphism, 
these are the linear orders that can be constructed from the empty 
linear order and a one point linear order  by the sum and 
product operations, reversal, and the operation $P \mapsto P^\omega$, where
\begin{eqnarray*}\label{}
P^\omega &=& \sum_{n \in \omega }  P^n. 
\end{eqnarray*}
This conjecture is supported by the fact 
that the scattered regular linear orders are exactly those that 
can be constructed from $0$, $1$ and $\omega$ by the sum and product
operations and reversal, cf. \cite{Heilbrunner}.  Thus, the
scattered regular linear orders have finite Hausdorff rank, but the
converse is false: the linear order
\begin{eqnarray*}
  \Z + 1 + \Z + 2 + \Z + \ldots + \Z + n + \Z + \ldots 
\end{eqnarray*}
where $\Z$ denotes the linear order of the negative and positive integers 
is not regular, but has finite Hausdorff rank.

After describing the scattered algebraic linear orders, the next task
could be to obtain a characterization of all algebraic linear orders.  We conjecture 
that these are exactly those linear orders that can be constructed from dense 
algebraic words by substituting a scattered algebraic linear order for each 
letter. (See below for the definition of an algebraic word.) Thus, the 
task can be reduced to the characterization of the dense algebraic words.

A hierarchy of recursion schemes was studied by
\cite{Damm,Damm82,Gallier,Ong,HagueOng}, and many others.
The schemes considered in this paper are on the first  
level of the hierarchy with regular schemes forming level $0$. 
In the light of the characterizations of the regular and algebraic 
ordinals, it is natural to conjecture that the ordinals definable 
on the $n$th level of the hierarchy are those less than 
$$\Uparrow(\omega,n+2) = \omega^{\omega^{\vdots^{ \omega}}}$$ where there are $n+2$ $\omega$'s altogether. 
In fact, every ordinal less than $ \Uparrow(\omega,n+2)$ is shown to 
be definable on the $n$th level in \cite{Braud}.

In ordinal analysis of logical theories, the strength of a theory is measured 
by ordinals. For example, the proof theoretic ordinal of Peano arithmetic is 
$\epsilon_0$. Here we have a similar phenomenon: we measure the strength 
of recursive definitions by ordinals, and we conjecture that the definable ordinals 
are exactly those less than $\epsilon_0$.

A generalization of the notion of ``finite word'' is obtained by considering 
labeled linear orders, where the labels are letters in some finite alphabet.
Thus, a linear order may be identified with a word on a one letter alphabet.
A \textbf{countable  
word} is word whose underlying linear order is countable. 
A morphism between words is a morphism between their respective underlying
linear orders that additionally preserves the labeling. Every countable 
word can be represented as the word determined by the frontier of a tree
where each leaf retains it label, cf. \cite{Courcelle78}.
Now an \textbf{algebraic word} (respectively \textbf{regular word}) 
is a word isomorphic to the frontier word of an algebraic (respectively, regular) 
tree. An ``operational'' characterization of the regular words was obtained in
\cite{Heilbrunner}, where it was shown that a nonempty word is regular
if and only if it can be constructed from single letter words by concatenation, 
$\omega$-power, the ``shuffle operations'' and reversal. (Note that concatenation corresponds 
to the sum operation on linear orders, and $\omega$-power to the operation $P \mapsto P \times \omega$.) 
Without the shuffle operations, exactly the nonempty scattered regular words can be generated,
and  the well-ordered regular words can be generated  by concatenation and $\omega$-power. 
It would be interesting to obtain operational characterizations of well-ordered, scattered,
and eventually, all algebraic words.  

Finally, we would like to mention an open problem. Suppose that 
a context-free language $L$ is well-ordered by the lexicographic order.
Is the order type of $(L,<_\ell)$ less than $\omega^{\omega^\omega}$?

\section{Acknowledgement}
The authors would like to thank the three referees whose suggestions
have resulted in an improved paper.  
   



\section*{Appendix}

This appendix is devoted to a  formal proof of the correctness of the translation 
given in Section~\ref{sec-translation}. 

Consider the system (\ref{a system}) whose least solution in $T_\Sigma^\omega(V)$ 
is $(T_1,T_2,\ldots, T_m)$. To this system, we can associate the 
\textit{tree grammar} $G_T$ whose productions are 
\begin{eqnarray*}
F_i(x_0,\ldots,x_{n_i-1}) &\to & t_i + \bot,\quad i =1,\ldots,m,
\end{eqnarray*}
where $\bot$ denotes the empty tree. The start symbol is $F_1$. Let $V$ denote the 
set of individual variables that occur in (\ref{a system}). 

Below we will also assume a new individual variable $z$ and write $t = t' \star t''$
for a tree $t$ over $\Sigma \cup \F$ possibly containing variables in $V$ 
if $t'$ is a tree with a \emph{single} leaf labeled $z$ and $t$ is obtained from $t'$ 
by replacing this leaf with a copy of $t''$. 

\begin{definition}
Suppose that $t,t'$ are finite trees in $T_\Sigma^\omega(V)$.
Then 
$t \implies t'$ in $G_T$ if $t$ can be written as $s \star F_i(s_0,\ldots,s_{n_i-1})$
for some trees $s$ and $s_0,\ldots,s_{n_i-1}$ 
such that  $t' = s \star t_i(s_0,\ldots,s_{n_i-1})$ or $t' = s\star \bot$. 
The relation $\dv$ is the reflexive transitive closure of 
$\implies$.  
\end{definition}

It is known, cf., \cite{CourcelleFund,Guessarian}, that for any $i = 1,\ldots,m$ and for any word $u$, 
$T_i(u)$ is defined if and only if there is some finite  tree $t$ \emph{in} $T_\Sigma^\omega(V)$ 
with $F_i(x_0,\ldots,x_{n_i-1}) \dv t$ such that $t(u)$ is defined, and in that 
case $T_i(u) = t(u)$. 

Thus, it suffices to prove that the grammar 
$G_L$ defined in 
Section~\ref{sec-translation} and the tree grammar $G_T$ are related 
as follows: 
\begin{lemma}
\label{lemma-A}
Let $i \in \{1,\ldots,m\}$ and  let $t$ be a finite tree  $T_\Sigma^\omega(V)$. 
Suppose that $F_i(x_0,\ldots,x_{n_i-1}) \dv t$. 
Then for every $u \in \dom(t)$ and $j \in [n_i]$, 
if $t(u) = x_j$ then $(F_i,j) \dv \widehat{u}$, and if $t(u) \in \Sigma_0 \cup \F$ then 
$F_i \dv \widehat{u}t(u)$. 
\end{lemma}

\begin{lemma}
\label{lemma-B}
Let $i \in \{1,\ldots,m\}$ and $j \in [n_i]$.
\begin{enumerate}
\item 
Suppose that $(F_i,j) \dv w$. Then there exist a finite tree $t$  in $T_\Sigma^\omega(V)$
and a word  $u\in \dom(t)$ such that 
$t(u) = x_j$, $w = \widehat{u}$ and $F_i(x_0,\ldots,x_{n_i-1}) \dv t$. 
\item Suppose that $F_i \dv w$. Then there exist $t$ and $u$ as above  with 
$t(u) \in \Sigma_0 \cup \F$, $w = \widehat{u}t(u)$ and $F_i(x_0,\ldots,x_{n_i-1}) 
\dv t$. 
\end{enumerate}
\end{lemma} 

{\sl Proof of Lemma~\ref{lemma-A}}. 
We argue by induction on the length of the derivation. When the length is $0$, 
$t = F_i(x_0,\ldots,x_{n_i-1})$. If $u = j$, for some $j \in [n_i]$, then 
$\widehat{u} = (F_i,j)$ and we clearly have $(F_i,j) \dv (F_i,j) = \widehat{u}$. 
If $u = \epsilon$ then $\widehat{u} = \epsilon$, $t(\epsilon) = F_i$ and we have
$F_i \dv F_i = \widehat{u}t(u)$.

In the induction step, assume that the length of the derivation is positive
and that our claim holds for all derivations of smaller length. 
Suppose that $t(u)$ is a variable $x_j$ or $t(u) \in \Sigma_0 \cup \F$. 
Let us 
write the derivation as 
$$F_i(x_0,\ldots,x_{n_i-1}) \dv t' \implies t$$
where in the last step we have $t' = s \star F_k(s_0,\ldots,s_{n_k-1})$
and $t = s\star t_k(s_0,\ldots,s_{n_k-1})$ or $t = s\star \bot$. 
In the second case, $u \in \dom(t')$, moreover $\widehat{u}$ 
in $t'$ is the same as $\widehat{u}$ in $t$, or as $\widehat{u}$ in $s$. Moreover, 
$t(u) = s(u) = t'(u)$. Thus, using the induction hypothesis, we obtain 
$(F_i,j) \dv \widehat{u}$ or 
$F_i  \dv \widehat{u}t'(u) = \widehat{u}t(u)$
according to whether $t(u) = x_j$ for some $j \in [n_i]$ or
$t(u) \in \Sigma_0 \cup \F$. 

Assume now that $t' = s\star F_k(s_0,\ldots,s_{n_k-1})$ and 
$t= s \star t_k(s_0,\ldots,s_{n_k-1})$. Let $v_0$ denote the 
unique word with $s(v_0) = z$. There are two cases. 
If $v_0$ is not a prefix of $u$, then we have that $u \in \dom(t') \cap \dom(s)$,
$\widehat{u}$ in $t$ is the same as $\widehat{u}$ in $t'$, 
and $t(u) = s(u) = t'(u)$. The proof is completed as before.  So let 
$v_0$ be a prefix of $u$. If there is some $w \in \dom(t_k)$ such that $u = v_0w$ and 
$t_k(w)$ is not an individual variable, then $\widehat{u}$ in $t$
is $\widehat{v}_0\widehat{w}$, where $\widehat{v}_0$ is computed in $s$ and 
$\widehat{w}$ is computed in $t_k$. Moreover, $t(u) = t_k(w) \in \Sigma_0 \cup \F$.
By the induction hypothesis we have $F_i \dv \widehat{v}_0F_k$, 
and by construction, $F_k \to \widehat{w} t_k(w)$ is a production. 
Thus, $F_i \dv \widehat{v}_0\widehat{w}t'(v_0w) = \widehat{v}_0\widehat{w}t_k(w) = \widehat{u}t(u)$. 

Suppose last that $u = v_0wv_1$, where $w \in \dom(t_k)$ with $t_k(w) = x_h$ 
for some $h \in [n_k]$ and $v_1 \in \dom(s_h)$. In that case $\widehat{u}$ in $t$ 
is $\widehat{v}_0\widehat{w}\widehat{v}_1$, where $\widehat{v}_0$ and $\widehat{w}$ are as before, and $\widehat{v}_1$ 
is computed in $s_h$. Moreover, $t(u) = s_h(v_1) = t'(v_0hv_1)$. Assume that 
$t(u)$ is the individual variable $x_j$. Then $(F_i,j) \dv \widehat{v}_0(F_k,h)\widehat{v}_1$ 
by the induction hypothesis, moreover, $(F_k,h) \to \widehat{w}$ is a production. 
We conclude that $(F_i,j) \dv \widehat{v}_0\widehat{w}\widehat{v}_1 = \widehat{u}$. 
Suppose now that $t(u) \in \Sigma_0 \cup \F$.
Then $F_i \dv \widehat{v}_0(F_k,h)\widehat{v}_1t'(v_0hv_1) =    
\widehat{v}_0(F_k,h)\widehat{v}_1s_h(v_1) 
\implies \widehat{v}_0\widehat{w}\widehat{v}_1 s_h(v_1) 
 = \widehat{u}t(u)$. \eop 

{\sl Proof of Lemma~\ref{lemma-B}}.
Suppose first that $(F_i,j) \dv w$. 
If the length of the derivation is $0$, our claim is trivial:
let $t = F_i(x_0,\ldots,x_{n_i-1})$, $u = j$.
We proceed by induction.  In the induction step,
we can write the derivation as $(F_i,j) \dv w_0(F_k,h)w_1 \implies 
w_0qw_1  = w$, where, by the induction hypothesis, there exist some $t'$ and $u_0,u_1$
with $t'(u_0hu_1) = x_j$, 
$w_0(F_k,h)w_1 = \widehat{u_0hu_1}$ 
in $t'$ and $F_i(x_0,\ldots,x_{n_i-1})  
\dv t'$. Since $(F_k,h) \to q$ 
is a production of $G_L$, there is some $p$ with $t_k(p) = x_h$
and $\widehat{p} = q$ in $t_k$. Clearly, we can write $t'$ as $t' = s\star F_k(s_0,\ldots,s_{n_k-1})$
where $s(u_0) = z$, so that $u_1 \in \dom(s_h)$ with $s_h(u_1) = x_j$. Now let 
$t = s \star t_k (s_0,\ldots,s_{n_k-1})$ and consider the 
 word $u = u_0pu_1$. We have that $t(u) = s_h(u_1) = t'(u_0hu_1) = x_j$ and 
$\widehat{u} = \widehat{u}_0\widehat{p} \widehat{u}_1 = w_0qw_1 = w$ in $t$. 

Suppose next that $F_i \dv w$. If the length of the derivation is $0$, then 
$w = F_i$ and we take $t = F_i(x_0,\ldots,x_{n_i-1})$ and $u = \epsilon$. 
Assume now that the length of the derivation is positive and that our claim
holds for shorter derivations. We can decompose the derivation either as 
$$F_i \dv w_0(F_k,h)w_1 \implies w_0qw_1 = w$$
or as 
$$F_i \dv w_0F_k \dv w_0w_1 = w.$$
The former case is similar to the previous one, so we only deal with the 
latter. In this case, by the induction hypothesis, there is some tree $t'$ 
and a word $u_0 \in \dom(t')$ with $t'(u_0) = F_k$, $w = \widehat{u}_0t'(u_0)$
and $F_i(x_0,\ldots,x_{n_i-1}) \dv t'$. Moreover, $F_k \to w_1$ 
is a production and thus $w_1 = \widehat{u}_1t_k(u_1)$ in $t_k$ for some $u_1$
with $t_k(u_1) \in \Sigma_0 \cup \F$.   
Since $t'(u_0) = F_k$, 
we can write $t'$ as $t' = s \star F_k(s_0,\ldots,s_{n_k-1})$,
where $s(u_0) = z$. Now let $t = s\star t_k(s_0,\ldots,s_{n_k -1})$
and let $u = u_0u_1$. We have that $F_i(x_0,\ldots,x_{n_i-1}) \dv t$ and 
$w = w_0w_1 = \widehat{u}t(u)$. Moreover, $t(u) = t_k(u_1) \in \Sigma_0 \cup \F$. \eop 

We can now prove the claim formulated in the proof of Theorem~\ref{alg to prefix}.
Suppose that $T_i(u) = x_j$ for some $i$, $1 \leq i \leq m$ and $j \in [n_i]$. 
Then $(F_i,j) \dv \widehat{u}$ by Lemma~\ref{lemma-A}.
And if $T_i(u) \in \Sigma_0$, then $(F_i,j) \dv \widehat{u}T_i(u)$. 
Conversely, if $(F_i,j) \dv w$ for some terminal word $w$, then 
by Lemma~\ref{lemma-B}, 
either $w = \widehat{u}$ for some $u$ with $T_i(u) = x_j$, or 
$w = \widehat{u}T_i(u)$ for some $u$ with $T_i(u) \in \Sigma_0$.  \eop 

\end{document}